\documentstyle[sprocl]{article}
\def\gtsim{\matrix{>\cr\noalign{\vskip-7pt}\sim\cr}}
\def\dsp{D^{*+}}

\def\pionm{{\vec p}_\pi}
\def\nc{{\cal N}_c}
\def\yo1{{{f_\pi}^2}}

\def\cpt{\chi -PT}

\def\bra#1{\langle{#1}\vert}
\def\ket#1{\vert{#1}\rangle}
\def\ha{ {\bar H}_a }
\def\hb{ H_b }
\def\sa{ {\bar S}_a }
\def\sb{ S_b }
\def\ta{ {\bar T}_a }
\def\tb{ T_b }
\def\as{ {\not \!\!{A}}_{ba} }

\def\gf{ \gamma _5}
\def\tbmu{ T_{b}^{\mu} }
\def\amu{ A_{\mu ba} }

\def\oneh{ {1\over 2} }
\def\threeh{ {3\over 2} }

\def\sss{\scriptscriptstyle}

\def\thefootnote{\fnsymbol{footnote}}

\begin{document}

\title{NOVEL ALGEBRAIC CONSEQUENCES OF CHIRAL SYMMETRY}
\author{Silas R.~Beane\,\footnote{Present Address: Physics Department, University of Maryland, College Park, MD 20742}}

\address{Department of Physics, Duke University, Durham, NC 27708-0305}

\maketitle\abstracts{ Phenomenologically motivated Lie-algebraic sum
rules determine the representations of {\it unbroken}
${SU(2)_L}\times{SU(2)_R}$ filled out by mesons containing a single
heavy quark, in the limit that the heavy quark mass goes to
infinity. This representation content determines the strong
single-pion transition amplitudes of all heavy meson states in the
chiral limit. Predictions are found to be in agreement with
experiment, and provide insight into the spectroscopy of heavy
mesons.}

\renewcommand{\thefootnote}{\alph{footnote}}
\setcounter{footnote}{0}

The dynamical consequences of broken chiral symmetry have proved
remarkably successful in explaining and predicting low-energy strong
interaction phenomena in a manner consistent with quantum
chromodynamics (QCD).  It is less well known that chiral symmetry has
additional consequences beyond those explored using chiral
perturbation theory ($\chi -PT$).  The nonperturbative QCD effects
which break chiral symmetry spontaneously also arrange hadrons into
{\it reducible} representations of the {\it unbroken} chiral
group\,\cite{alg}.  In principle any number of hadrons can participate
in a given reducible chiral multiplet. Moreover, the angles which mix
the various irreducible representations are not fixed by any QCD
symmetry. One might therefore conclude that these algebraic
consequences of chiral symmetry cannot yield much predictive
power. However, it turns out that several phenomenologically inspired
Lie-algebraic sum rules severely constrain both the permissible
particle content and the mixing angles of these
representations\,\cite{alg}$^{\!,\,}$\,\cite{mended}$^{\!,\,}$\,\cite{dirac}.

Here we concentrate on systems of mesons that transform in
the $I=\oneh$ representation of the diagonal isospin subgroup of
${SU(2)_L}\times{SU(2)_R}$. We assume that pion scattering processes
involving these mesons are determined by the sum of all chiral {\it
tree graphs}.  This is certainly true in the large-$\nc$
limit\,\cite{hooft}.  However, if the mass splitting between any two
members of a given chiral multiplet is small compared to
$\Lambda_\chi$, then pions can be considered soft and restriction to
chiral tree graphs is automatic when one works to leading order in
$\chi -PT$. We take this point of view since it allows us to express
pion scattering amplitudes in terms of the undetermined constants that
appear in $\chi -PT$.  The algebraic content of chiral symmetry allows
one to prove that, for a given helicity, only pairs of $I=\oneh$
mesons communicate by single-pion emission and absorption\,\cite{dirac}.
We consider the consequences of this result for mesons that carry a
single heavy quark. We show that constraints on the heavy meson mass
matrix that arise from QCD in the heavy quark expansion determine the
reducible chiral multiplets of unbroken ${SU(2)_L}\times{SU(2)_R} $
filled out by the infinite tower of heavy meson doublets. This
representation content predicts all strong single-pion transitions of
these states. For example, the transition amplitude for the process
${P^*}\rightarrow P\pi$ is found to vanish identically.  These purely
group-theoretical predictions are consistent with observation, and
yield interesting insight into the spectroscopy of heavy mesons.

First we review some essential formalism. In practice, the algebraic
consequences of chiral symmetry arise from a need for cancellations
among chiral tree graphs involving pions ---constructed from the most
general chiral invariant operators\,\cite{alg}. In the chiral limit, a
generalized Adler-Weisberger sum rule can be expressed as

\begin{equation}
{\lbrack {{X_{i}}},\,{{X_{j}}}\rbrack}
=i\epsilon _{ijk}{T_{k}},
\end{equation}
where ${T_i}$ is the isospin matrix, and
${X_i}$ is an axial-vector coupling matrix, related to the
matrix element of the process ${\alpha (p,\lambda )}\rightarrow{\beta
(p',\lambda')}+{\pi}{(q,i)}$ in any frame in which the momenta are
{\it collinear}.
Here $\alpha$ and $\beta$ are arbitrary single-hadron states and 
$\lambda$ is helicity ---which is conserved in the collinear
frame. Together with Eq.~(1), the defining relations, $\lbrack
T_{i},\,T_{j}\rbrack=i\epsilon_{ijk}T_{k}$ and $\lbrack
T_{i},\,{{X_{j}}}\rbrack= i\epsilon _{ijk}{{X_{k}}}$, close the chiral
algebra and we see that for each helicity, $\lambda$, hadrons fall
into representations of ${SU(2)_L}\times{SU(2)_R} $, {\it in spite of
the fact that the group is spontaneously broken}.  However, ${X_i}$
does not commute with the mass-squared matrix, ${\hat m}^2$, and
therefore, in general, these representations are reducible.  More
information about the mass-squared matrix follows from the sum rule

\begin{equation}
{\lbrack {X_{j}},\,\lbrack {X_{i}},\,{ m^{2}}\rbrack\rbrack}\propto
\delta_{ij}.
\end{equation} 
For a system of $I=\oneh$ mesons, with no single-pion transitions to
states of higher isospin, this sum rule requires no assumption beyond
Eq.~(1).  This sum rule implies that the hadronic mass-squared matrix
is the sum of a chiral invariant and the fourth component of a chiral
4-vector; {\it i.e.}  ${{\hat m}^2}$=${{\hat m}_0^2}$+${{\hat
m}_4^2}$.  The only representations of
${SU(2)_L}\times{SU(2)_R} $ that contain only a single $I={\oneh}$
representation of the diagonal isospin subgroup are $(0,{{\oneh}})$
and $({{\oneh}},0)$.  So, in general, $I=\oneh$ states of definite
helicity are linear combinations of any number of these irreducible
representations with undetermined coefficients\,\cite{dirac}.  Mass
splitting can only occur as a consequence of mixing between these
representations since ${\hat m}^2$ is a sum of $(0,0)$ and
$({{\oneh}},{{\oneh}})$ contributions.  In a basis in which all linear
combinations of $(0,{{\oneh}})$, and $({{\oneh}},0)$ irreducible
representations appear in that order, the mass-squared matrix takes
the supermatrix form

\begin{equation}{{\hat m}^2}=\left(\matrix{{\hat A}&0\cr
                      0&{\hat B}\cr}\right) 
 +\left(\matrix{0&{\hat G}\cr
      {{\hat G}^\dagger}&0\cr}\right). 
\end{equation} 
The case of zero-helicity is of special interest. The $SU(2)\times
SU(2)$ algebra has an {\it endomorphism}, $\Pi$: $\Pi{X_i}\Pi=-{X_i}$,
and $\Pi{T_i}\Pi={T_i}$\,\cite{mended}. The eigenvalues of $\Pi$ are
{\it normality}, ${\eta _\alpha}\equiv {P_\alpha}{{(-)}^{J_\alpha}}$,
where ${P_\alpha}$ and $J_\alpha$ are the intrinsic parity and spin of
$\alpha$, respectively. Only zero-helicity states of opposite
normality communicate by single-pion emission and absorption.  The
effect of the normality operator is to change $(0,{{\oneh}})$
representations into $({{\oneh}},0)$ representations and vice versa.
Since $\Pi$ commutes with ${\hat m}^2$, it follows that ${\hat
A}={\hat B}$ and ${\hat G}={{\hat G}^\dagger}$. Eigenstates with $\eta
=(\pm )$ are eigenvectors of ${\hat A}\pm {\hat G}$.  The matrix
element of ${X_i}$ between two arbitrary $I=\oneh$ states, $\alpha$
and $\beta$, of opposite normality can be written as
$\bra{\beta}{X_i}\ket{\alpha}\equiv{T_i}g_{\sss \beta\alpha\pi}$,
where ${T_i}={\tau _i}/2$ and the ${\tau _i}$ are the Pauli matrices.
By building physical particle states out of fundamental states of
definite normality, it is straightforward to show that $|g_{\sss
\beta\alpha\pi}|\leq 1$; an {\it exact} consequence of chiral
symmetry\,\cite{silas}.

We assume that no $I=0$ Regge trajectories contribute to transitions
in which there are different particles in the initial and final
state\,\cite{alg}$^{\!,\,}$\,\cite{mended}.  In other words, we assume
that scattering becomes purely elastic at high energies.  Phenomena
suggest that this should be a very good approximation.  For example,
the cross sections for the processes $\pi +N \rightarrow {a_{\sss
1}}(1260)+N$ and $N+N\rightarrow {N^*}(1440)+N$ are less than $10\%$
of those for $\pi +N\rightarrow\pi +N$ and $N+N\rightarrow N+N$,
respectively\,\cite{anti}$^{\!,\,}$\,\cite{biel}. Although this can be
understood on the basis of simple ``diffraction'' arguments, it is
certainly not clear ---from the QCD point of view--- why this is the
case. Here we treat this constraint as experimental input.  This
assumption leads to a {\it superconvergence} relation, which can be
expressed in Lie-algebraic form
as\,\cite{alg}$^{\!,\,}$\,\cite{mended}

\begin{equation}{\lbrack { m^{2}},\,\lbrack {X_{i}},\,
\lbrack {X_{j}},\, {m^{2}}\rbrack\rbrack\rbrack}=0. 
\end{equation} 
This sum rule is simply the statement that ${\hat m}_0^2$ and ${\hat
m}_4^2$ commute, or using Eq.~(5), $\bra{\alpha}{{\hat
m}_4^2}\ket{\beta}=0$ when $\alpha\neq\beta$. We can immediately
extract the general consequences of this sum rule for the helicity
zero states of the $I=\oneh $ mesons.  Below we specialize a general
theorem, proved in Ref.~3, to zero helicity.

As shown above, physical eigenstates of $\eta=(\pm )$ are
eigenstates of ${\hat A}\pm{\hat G}$.
Eq.~(5) implies that ${\hat A}$ and ${\hat G}$ commute. 
Suppose that the vector ${\vec a}$ represents a physical state in the
$\eta =(+)$ basis. Since ${\hat A}$ and ${\hat G}$ commute, ${\vec a}$
is a simultaneous eigenvector of ${\hat A}$ and ${\hat G}$, say with
eigenvalues $\mu ^2$ and $\Delta $, respectively. Similarly, suppose
that the vector ${\vec b}$ represents a physical state in the $\eta
=(-)$ basis; ${\vec b}$ is also a simultaneous eigenvector of ${\hat
A}$ and ${\hat G}$. There are then two possibilities: ${\vec a}$ and
${\vec b}$ have different eigenvalues in which case ${\vec
a}\cdot{\vec b}=0$, or ${\vec a}$ and ${\vec b}$ have the same
eigenvalues in which case $|{\vec a}\cdot{\vec b}|=1$.  It follows
that only pairs of states with masses ${\mu ^2}\pm\Delta $ communicate
by single-pion emission and absorption.  In terms of coupling
constants, $|g_{\sss \beta\alpha\pi}|= 1$ if $\alpha$ and $\beta$ are
paired, or $g_{\sss \beta\alpha\pi}= 0$, otherwise.  This result is
completely general and applies to all $\lambda =0$ states of the
$I=\oneh$ mesons.  The interactions of heavy mesons with
pions are also constrained by heavy quark symmetry.

Heavy hadron $\chi -PT$ provides an expansion in powers of momenta
divided by ${\Lambda_\chi}\sim{m_\rho}$, and in powers of
$\Lambda_{QCD}$ divided by the heavy hadron mass\,\cite{wise}.  Mesons
containing a heavy quark can be classified by the spin ($s_{\ell}$)
and the parity ($\pi _{\ell}$) of the light
quark\,\cite{isgur}. Consequently, heavy mesons fall into degenerate
doublets labelled by ${s_{\pm}^\pi}=(s_{\ell}\pm{{\oneh}})^{\pi
_{\ell}}$.  The ground state mesons have ${s_{\ell}}={\oneh}$ and
${\pi _{\ell}}=(-)$ and are denoted $P$ ($0^-$) and $P^*$
($1^-$)\,\footnote{ $P$ can be a $D$ or a $B$ meson, and the ``$*$''
superscript indicates positive normality ($\eta =(+)$), since, by
convention, ``$*$'' is assigned to states in the spin-parity series
$J^P= {0^+},{1^-},{2^+},\dots$.}. The first excited states have
${s_{\ell}}={\oneh}$ and ${\pi _{\ell}}=(+)$, and are denoted $P_0^*$
($0^+$) and $P_1'$ ($1^+$).  At the next level we have
${s_{\ell}}={\threeh}$ and ${\pi _{\ell}}=(+)$, corresponding to $P_1$
($1^+$) and $P_2^*$ ($2^+$).  The $\oneh ^-$, $\oneh ^+$, and $\threeh
^+$ doublets can be assembled into the ``superfields'' ${H_a}$,
${S_a}$, and ${T_a}$, respectively.  The pion transitions within the
heavy meson doublets are contained in the
operators\,\cite{wise}$^{\!,\,}$\,\cite{falk}


\begin{equation}
g\; {\rm Tr}[\ha\hb\as\gf]
+{g^\prime}\; {\rm Tr}[\sa\sb\as\gf]
+{g^{\prime\prime}}\; {\rm Tr}[\ta\tb\as\gf] 
\end{equation}

\begin{equation}
{f^\prime}\; {\rm Tr}[\sa\tbmu\amu\gf]
             +{h}\; {\rm Tr}[\ha\sb\as\gf]+h.c. ,
\end{equation}
where ${A_{ba}}$ is the usual axial-vector Goldstone boson matrix.

We are now in a position to consider the joint consequences of
unbroken ${SU(2)_L}\times{SU(2)_R} $ and heavy quark symmetry.  We first
relate the coupling constants that appear in the effective lagrangian
to matrix elements of the axial-vector matrix, ${X_i}$:

\begin{eqnarray}
&&\qquad
\bra{P}{X_i}\ket{P^*}={g}{T_i}\quad\bra{P_1'}{X_i}\ket{P_0^*}={g^\prime}{T_i}
\quad\bra{P_1}{X_i}\ket{P_2^*}={g^{\prime\prime}}{T_i}\nonumber \\ 
&&\bra{P_1}{X_i}\ket{P_0^*}=\bra{P_1'}{X_i}\ket{P_2^*}={f^{\prime}}{T_i}\quad
\bra{P}{X_i}\ket{P_0^*}=\bra{P_1'}{X_i}\ket{P^*}={h}{T_i}.
\end{eqnarray}
The overall phases have been fixed by convention as we will be
concerned only with absolute values.  In making this identification,
we are clearly working to leading order in heavy hadron chiral
perturbation theory; hence, the identification in Eq.~(7) holds
for each helicity, up to an overall phase.  The orbital angular
momentum of a single-pion transition is subject to the selection rule
${P_\alpha}{P_\beta}=-{(-)^{\ell}}$ where ${P_\alpha}$ is the
intrinsic parity of $\alpha$.  If $\alpha$ and $\beta$ are nearly
degenerate ---a necessary condition for the applicability of $\cpt$---
then the contribution to the transition of the $\ell$th partial wave
will be of order $({M_\alpha^2}-{M_\beta^2})^{\ell}$ and so one can
neglect higher partial waves\,\cite{alg}.  Hence to leading order in
heavy hadron $\cpt$ the single-pion transitions within the heavy meson
doublets are P-wave and between adjacent doublets of opposite parity
are S-wave (unless there are further selection rules).  These
transition matrix elements are bounded as a consequence of the
participation of heavy mesons in multiplets of unbroken chiral
symmetry: $g$, ${h}$, ${g'}$, ${f^{\prime}}$ and ${g^{\prime\prime}}$
must all take values between $-1$ and $1$\,\cite{chow}.

Consider the low-lying heavy meson doublets.  The general theorem
deduced from the three chiral commutation relations allows two
scenarios consistent with heavy quark symmetry: (i) $P$ is paired with
$P_0^*$, and $P^*$ is paired with $P_1'$. This yields $|{h}|=1$,
$g={g'}={f^{\prime}}=0$.  (ii) $P$ is paired with $P^*$.  This yields
$|g|=1$ and ${h}=0$. Of course, here we assume that only adjacent
heavy meson states participate in a given chiral multiplet. One might
suppose that since $P_0^*$ and $P_1'$ are unobserved, $P$ could be
paired with $P_{\sss 2}^*$. However, this transition is forbidden by
heavy quark symmetry\,\cite{isgur}. So in the absence of the $\oneh ^+$
doublet, case (ii) would have to be realized. 

It is straightforward to show that case (ii) is inconsistent with QCD
in the heavy quark limit.  Heavy quark symmetry constrains the heavy
hadron mass matrix\,\cite{aglietti}$^{\!,\,}$\,\cite{wise2}.  These
constraints are easily translated to the mass-squared matrix. In the
heavy quark limit, the combination $3{M_{P^*}^2}+{M_P^2}$ is
independent of the mass-squared difference:
${M_{P^*}^2}-{M_P^2}$. This is a simple consequence of the fact that
the leading mass splitting between $P$ and $P^*$ arises from the
coupling of the spin of the light quark to the spin of the heavy
quark. If $P$ and $P^*$ are paired in the sense of case (ii), then we
know from the general theorem derived above that their squared-masses
can be written as ${\mu ^2}\pm\Delta $, where $\Delta$ $(\neq 0)$ is a
diagonal element of the matrix ${\hat m}_4^2$.  However, it is then
clearly impossible that $3{M_{P^*}^2}+{M_P^2}$ be independent of
$\Delta$ in the heavy quark limit. Therefore, case (ii) is
inconsistent with QCD, whereas case (i) easily accomodates the QCD
constraint. Since no two members of a heavy meson doublet have the
same spin, there exist similar constraints on the mass-squared matrix
of any arbitrary heavy meson doublet, and so the chiral pairing of the
two lowest lying doublets should be realized by all heavy meson {\it
quartets} labelled by the light quark spin.  Therefore, in the heavy
quark limit, the algebraic constraints uniquely
determine the single-pion transition amplitudes of {\it all} heavy
meson states.  This is our main result.  As regards the lowest lying
doublets, we conclude that $g={g'}={g^{\prime\prime}}={f^{\prime}}=0$
and $|{h}|=1$\,\cite{bardeen}.
What reducible representations of ${SU(2)_L}\times{SU(2)_R} $ does this
solution correspond to?  It is straightforward to show that the
zero-helicity states of heavy mesons of a given light quark spin fall
into quartets composed of reducible doublets of opposite
normality\,\cite{silas}. Other-helicity states fall into reducible
doublets. 

There is not a great deal of experimental data available for the
strong pion transitions of the heavy meson states.  Here we will see
how our predictions compare with existing experimental results.  Note
that the solution to the chiral commutation relations is not intended
to give a detailed fit to the heavy meson spectrum; in fact, many
transitions forbidden by our selection rule occur in nature.  Rather
the purpose of this work is to understand general features using
symmetries of QCD.  Much effort has centered on determining the
${P^*}\rightarrow P\pi$ transition amplitude.  The kinematically
allowed decay ${D^*}\rightarrow D\pi$ is determined by $g$:

\begin{equation}
\Gamma (\dsp\rightarrow{D^0}{\pi ^+})=
{{g^2}\over{12\pi\yo1}}{|\pionm |}^3 .
\end{equation}
In our convention, ${f_\pi}=93\;{\rm MeV}$.  There is an experimental
upper limit on the $D^*$ width\,\cite{decaydata}: ${{g}^2}<0.5$. The
radiative $D^*$ decays offer an indirect method of determining $g$,
and lead to the lower bound ${{g}^2}{\gtsim}0.1$\,\cite{amundson}.

There are two sorts of chiral symmetry breaking effects that we have
to consider.  In the decay width formula for ${D^*}\rightarrow D\pi$
we have used physical pions to compute kinematical factors, and yet
the coupling constant $g$ was evaluated at zero pion mass; a non-zero
pion mass interferes with the counting of powers of energy which is
essential in deriving the Lie-algebraic sum rules\,\cite{alg}.  We know
of no way of accounting for this small effect in a systematic fashion.
The second type of breaking arises from chiral symmetry breaking
operators in the effective
lagrangian\,\cite{goity}.  These effects lead to
an effective coupling constant of the form ${g_{eff}}=g\lbrace
1+O({m_q})\rbrace$, where the corrections arise from non-analytic (in
$m_q$) one loop graphs constructed from the leading order
operators. This sort of correction is generically large\,\cite{goity};
{\it i.e.}, a $20\%$ effect.  However, these corrections are {\it
weighted} by $g$, and so vanish along with the axial-vector source.
Evidently, the prediction $g=0$ is not subject to explicit chiral
symmetry breaking effects of this type.  Therefore, deviations of $g$
from $0$ should be due entirely to chiral symmetry breaking effects of
the kinematic type discussed above.  In other words, our solution
unambiguously predicts that the transition amplitude for the process
${P^*}\rightarrow P\pi$ should be very close to $0$, and therefore the
decay width for the process ${D^*}\rightarrow D\pi$ should be very
close to the experimental lower bound implied by the radiative decays.
The issue of $1/M$ corrections is discussed in Ref. 5.

The $\oneh ^+$ states have not been observed in the $D$ and $B$ meson
systems.  Of course these states are expected to exist.  If one takes
${M_{D_1'}}={M_{D_0^*}}=2.4\; {\rm GeV}$ as suggested by the quark
model\,\cite{quarkm} one finds $\Gamma ({D_0^*}\rightarrow
D{\pi^-})={|{h}|^2}\;[980\;{\rm MeV}]$ and $\Gamma ({D_1'}\rightarrow
{D^*}{\pi^-})={|{h}|^2}\; [400\;{\rm MeV}]$ and so {\it a priori} it
is not surprising that these states are unobserved\,\cite{falk}. These
results are quite sensitive to the choice of the heavy meson masses,
and yet it is gratifying that our general solution, $|{h}| =1$,
implies that these widths should take the {\it maximum} values allowed
by unbroken chiral symmetry.  The alternative solution, ${h}=0$, would
unambiguously predict that these states are narrow.

\subsection*{Acknowledgements}

The author thanks the organizers of the Workshop on "QCD: Collisions,
Confinement, and Chaos", Paris, France, June 96. This work was
supported by the U.S. Department of Energy (Grant DE-FG05-90ER40592)
at Duke (DUKE-TH-95-100).


\begin{thebibliography}{}

\bibitem{alg}  S.~Weinberg, {\sl Phys. Rev. {\bf 177}} (1969) 2604. 

\bibitem{mended}  S.~Weinberg, {\sl Phys. Rev. Lett. {\bf 65}} (1990) 1177. 

\bibitem{dirac} S.~Weinberg, {\sl Phys. Rev. Lett. {\bf 65}} (1990) 1181. 

\bibitem{hooft}  G.'t Hooft, {\sl Nucl. Phys. {\bf B72}} (1974) 461;
E.~Witten, {\sl Nucl. Phys. {\bf B160}} (1979) 57. 

\bibitem{silas} S.R.~Beane, {\sl DUKE-TH-95-98}, hep-ph/9512228.

\bibitem{anti} Yu.M.~Antipov {\it et al},
                  {\sl Nucl. Phys. {\bf B63}} (1973) 141. 

\bibitem{biel} J.~Biel {\it et al}, 
{\sl Phys. Rev. Lett. {\bf 36}} (1976) 504; 
H.~de Kerret {\it et al},
{\sl Phys. Lett. {\bf B63}} (1976) 477; 
J.~Biel {\it et al},
{\sl Phys. Lett. {\bf B65}} (1976) 291. 

\bibitem{wise} M.B.~Wise,  {\sl Phys. Rev. {\bf D45}} (1992) R2188. 

\bibitem{isgur} N.~Isgur and M.B.~Wise,  
                 {\sl Phys. Rev. Lett. {\bf 66}} (1991) 1130. 

\bibitem{falk} A.F.~Falk and M.~Luke,
{\sl Phys. Lett. {\bf B292}} (1992) 119; 
U.~Kilian, J.G.~K{\" o}rner, and D.~Pirjol, 
{\sl Phys. Lett. {\bf B288}} (1992) 119. 

\bibitem{chow} See also,
               C-K.~Chow and D.~Pirjol,{\sl CLNS-95/1377}, hep-ph/9512242.

\bibitem{aglietti} U.~Aglietti,
                 {\sl Phys. Lett. {\bf B281}} (1992) 341. 

\bibitem{wise2} M.B.~Wise, {\sl CALT-68-1901}, hep-ph/9311212.

\bibitem{bardeen} See also, W.A.~Bardeen and C.T.~Hill
                  {\sl Phys. Rev. {\bf D49}} (1994) 409. 

\bibitem{decaydata} ACCMOR Collab. S.~Barlag {\it et al}
                 {\sl Phys. Lett. {\bf B278}} (1992) 480. 

\bibitem{amundson} J.F.~Amundson {\it et al},
{\sl Phys. Lett. {\bf B296}} (1992) 415 
P.~Cho and H.~Georgi, {\sl Phys. Lett. {\bf B296}} (1992) 408. 

\bibitem{goity} J.L.~Goity,
{\sl Phys. Rev. {\bf D46}} (1992) 3929; 
H-Y.~Cheng {\it et al},
{\sl Phys. Rev. {\bf D49}} (1994) 5857. 

\bibitem{quarkm} S.~Godfrey and N.~Isgur,
{\sl Phys. Rev. {\bf D32}} (1985) 189; 
S.~Godfrey and R.~Kokoski,
{\sl Phys. Rev. {\bf D43}} (1991) 1679. 

\end{thebibliography}
\end{document}